\documentclass[12pt,preprint]{aastex}

\slugcomment{submitted to the PASP}

\shorttitle{FUV Spectroscopy of Three Nova-Like Variables
}
\shortauthors{}

\begin{document}

\title{Far Ultraviolet Spectroscopy of Three Long Period Nova-Like Variables
}

\author{Alexandra C. Bisol, 
Patrick Godon\altaffilmark{1}, \& Edward M. Sion
}
\affil{Astronomy \& Astrophysics, Villanova University, \\
800 Lancaster Avenue, Villanova, PA 19085, USA}
\email{abisol01@villanova.edu, patrick.godon@villanova.edu, 
edward.sion@villanova.edu}
\altaffiltext{1}{Visiting at the Henry A. Rowland Department of Physics
\& Astronomy, the Johns Hopkins University, Baltimore, MD 21218, USA} 

\begin{abstract}

We have selected three nova-like variables at the long period extreme of 
nova-like orbital periods: V363 Aur, RZ Gru and AC Cnc, all with IUE 
archival far ultraviolet spectra. All are UX UMa type nova-like variables and all have 
$P_{orb} > 7$h. V363 Aur is a bona fide SW Sex star, and AC Cnc is a probable one, while RZ Gru has not proven to be a member of the SW Sex subclass.
We have carried out the first synthetic spectral analysis of far ultraviolet spectra of the three systems using 
state-of-the-art models both of accretion disks and white dwarf photospheres.
We find that the FUV spectral energy distribution of both V363 Aur and RZ Gru 
are in agreement with optically thick steady state accretion disk models in which the luminous disk accounts for 100\% of the FUV light. 
We present accretion rates and model-derived distances for V363 Aur and RZ Gru. For AC Cnc, we find that a hot accreting white dwarf 
accounts for $\sim$~ 60\% of the FUV light with an accretion disk providing the rest. We compare our accretion rates and model-derived 
distances with estimates in the literature. 

\end{abstract}


\section{Introduction}

Nova-like variables are a subset of Cataclysmic variables (CVs), short-period binaries in which a late-type, Roche-lobe-filling
main-sequence dwarf transfers gas through an accretion disk onto a rotating, accretion-heated white dwarf (WD). The spectra of nova-like variables resemble those of classical novae (CNe) that have settled back to quiescence. However, they have never had a recorded CN outburst or any outburst. Hence
their evolutionary status is unclear: they could be close to having their next CN explosion, or they may have had an unrecorded explosion in the past, possibly hundreds or thousands of years ago. Or, quite possibly, they may not even have CN outbursts at all, in which case their WDs would be steadily increasing in mass to become the elusive progenitors of Type Ia supernovae.

A subset of nova-likes are those which remain in a high brightness state as though they are dwarf novae stuck in outburst. These nova-likes are part of the UX UMa subset. Complicating the picture of nova-likes further, are the SW Sextantis subset. They  display a multitude of observational characteristics: orbital periods between 3 and 4 hours;
one third of the systems are non-eclipsing while two-thirds show deep eclipses of the WD by the secondary (thus requiring high inclination angles); single-peaked emission lines despite the high inclination, and high excitation spectral features including He II (4686) emission and strong Balmer emission on a blue continuum; high velocity emission S-waves with maximum blueshift near phase $\sim$  0.5; delay of emission line radial velocities relative to the motion of the WD; and central absorption dips in the emission lines around phase $\sim$ 0.4 - 0.7 (Rodriguez-Gil, Schmidtobreick \& Gaensicke 2007; Hoard et al.2003).  
The white dwarfs in many if not all of these systems are suspected of 
being magnetic (Rodriguez-Gil et al. 2007) although this hypothesis remains unproven. Since SW Sex stars as well as most of the UX UMa systems are found near the upper boundary of the two to three hour period gap, a much better understanding of them is of critical importance to understanding CV evolution as they enter the period gap (Rodriguez-Gil et al.2007), if indeed they even do enter the gap, since evolution across the gap has not yet been definitively proven.
  
Recent systematic studies of a larger number of nova-like systems have been carried out by Puebla et al (2007), Ballouz and Sion (2009), Zellem et al. (2009) and Misuzawa et al.(2010) with the aim of modeling their FUV spectra and comparing accretion rates among different subgroups of CVs as a function of orbital period. For this paper, we have selected three nova-like variables at the long period extreme of nova-like orbital periods: V363 Aur, RZ Gru and AC Cnc,  all of which are UX UMa type nova-like variables with orbital periods longer than 7 hours, well beyond the 3 to 4 hour range where most nova-likes are found.  
Lanning 10 (hereafter V363 Aur) is a bona fide SW Sex star, and AC Cnc is a probable one, while RZ Gru has not been proven to be an SW Sex star. All three are part of the UX UMa subset. In this paper, our approach is to utilize grids of {\it composite} models (accretion disk plus white dwarf) for the first time on the IUE spectra of these three systems. The models are the same as used on dwarf nova in outburst (Hamilton et al. 2007). Among the key questions we seek to answer are: What is the accretion rate? Is this rate of accretion consistent with the structure of a steady-state disk model? If the white dwarf is exposed, then how hot is it  and how much flux does it contribute to the FUV?  How much flux does the accretion disk contribute to the FUV? Given that all three systems have $P_{orb} >  7$ hours and well beyond the observed concentration of nova-like system orbital periods between three and four hours, it is interesting to explore any differences these three long period systems might exhibit. 

The known orbital and physical parameters from the literature on the three systems are presented in Table 1 along with the references. We list for these three systems, by row: (1) the NL subtype(s); (2) average apparent V-magnitude; (3) the orbital period in hours; (4) the interstellar reddening [E($\bv$)]; (5) mass of the white dwarf in solar masses; (6) mass of the red dwarf in solar masses; (7) the orbital inclination i and; (8) the distance in parsecs.

\begin{deluxetable}{lrrc}
\tabletypesize{\scriptsize}
\tablecaption{Parameters}
\tablecolumns{4}
\tablewidth{0pt}
\tablehead{
\colhead{} & \colhead{V363 Aur} & \colhead{RZ Gruis}   & \colhead{AC Cancri}
}
\startdata

Subtype     & UX,SW               & UX                    & UX,SW \\
V           & 14.2                & 12.3                  & 13.5 \\
Period      & $7h 42m^{a}$        & $8h 38m - 9h 48m^{c}$ & $7.21^{i}$ \\
E(B-V)      & $0.3^{a}$           & $0.03^{b}$            & $0.055^{j}$ \\
$M_{wd}$    & $0.86 M_{\sun}^{h}$ & \nodata       & $0.76 \pm 0.03 M_{\sun}^{i}$ \\
$M_{rd}$    & $0.77 M_{\sun}^{d}$ & \nodata    & $1.02 \pm 0.14 M_{\sun}^{i}$ \\
Inclination & $73\,^{\circ}{\rm}^{ f}$&$61\,^{\circ}{\rm}^{ g}$& $75.6\pm0.03\,^{\circ}{\rm}^{ i}$ \\
Distance    & $530-1000 pc^{f}$   & $440 \pm40 pc^{e}$    & $550^{i}$ \\
\enddata
\tablerefs {a. Szkody et al. 1981, b. Bruch et al. 1993, c. Tappert et al. 1998, d. Schegel et al. 1986, e. Stickland et al. 1984, f. Rutten et al. 1992, g. Kelly et al. 1981, h. Honeycutt et al. 1986, i. Thoroughgood et al.2004, j. Galex Archive}
\end{deluxetable}

The listed values of white dwarf mass, orbital inclination, and the reddening E(B-V) were adopted as initial values in the model fitting. The wide range of published distance determinations for V363 Aur as well as the single distance estimates published for RZ Gru and AC Cnc allow a useful comparison with the model-derived distances from this work. Unfortunately, due to the long orbital periods of V363  Aur,  RZ Gru and AC Cnc, we could not use  
Knigge's (2006) distance method which is based upon 2MASS JHK photometry 
and a semi-empirical CV donor sequence which extends only to $P_{orb} = 6$h. 

\section{IUE Archival Observations}

We extracted the archival spectra for each system from the MAST archive. 
An observing log is given in Table 2 for the IUE SWP spectra selected for our analysis. For the three systems we list by row: (1) the SWP \#; (2) the exposure time in seconds; (3) the spectral dispersion; (4) the aperture size; (5) the date and time of the start of the exposure; (6) the maximum continuum counts of each exposure; and (7) the background counts of each exposure.

\begin{deluxetable}{lrrc}
\tabletypesize{\scriptsize}
\tablecaption{IUE Observation Log}
\tablecolumns{4}
\tablewidth{0pt}
\tablehead{
\colhead{}            & \colhead{V363 Aur} & \colhead{RZ Gruis} & \colhead{AC Cnc} 
}
\startdata
Spectra               & SWP35335            & SWP18138            & SWP18734  \\
Exposure Time         & 9000s               & 2100s               & 6000s    \\
Dispersion            & Low                 & Low                 & Low  \\
Aperature             & Large               & Large               & large  \\
Obervation Start Time & 1985-02-28 04:30:18 & 1982-09-27 21:15:53 & 1982-12-05 03:47:36  \\
Max Continuum         & 75                  & 150                 & 51  \\
Background Count      & 32                  & 21                  & 27  \\
\enddata
\end{deluxetable}

All three systems were in their normal high brightness state when the IUE spectra were obtained. We used IUERDAF software to process the spectra and the IDL routine UNRED to de-redden the spectra. The IUE SWP spectrum of V363 Aur reveals emission lines the strongest of which are due to C IV and He II plus weaker emission features due to S I, C I, Al II, and C III. The only absorption
features evident in the spectrum are Si III + O I (1300) and C II (1335). These features are labeled in Fig.1 which is discussed in section 3 below.
The spectrum of RZ Gru strongly resembles a UX UMa type nova-like system with strong P Cygni profiles at N V, SiIV and C IV revealing the presence of
a hot, fast outflowing wind. These features are displayed in Fig.2 which is discussed in Section 3 below.     

\section{Synthetic Spectral Fitting Models}

We adopted model accretion disks from the optically thick disk model grid 
of Wade \& Hubeny (1998). In these accretion disk models, the innermost disk radius, R$_{in}$, is fixed at a fractional white dwarf radius
of $x = R_{in}/R_{wd} = 1.05$. The outermost disk radius, R$_{out}$, was chosen so that T$_{eff}(R_{out})$ is near 10,000K since disk annuli beyond this point, which are cooler zones with larger radii, would provide only a very small contribution to the mid and far UV disk flux, particularly the SWP FUV bandpass. The mass transfer rate is assumed to be the same for all radii.
Thus, the run of disk temperature with radius is taken to be:

\begin{equation}
T_{eff}(r)= T_{s}x^{-3/4} (1 - x^{-1/2})^{1/4}
\end{equation}

where  $x = r/R_{wd}$
and $\sigma T_{s}^{4} =  3 G M_{wd}\dot{M}/8\pi R_{wd}^{3}$

Limb darkening of the disk is fully taken into account in the manner described by Diaz et al. (1996) involving the Eddington-Barbier relation,
the increase of kinetic temperature with depth in the disk, and the wavelength and temperature dependence of the Planck function. The boundary layer contribution to the model flux is not included. However, the boundary layer is expected to contribute primarily in the extreme ultraviolet below the Lyman limit. 

Theoretical, high gravity, solar composition photospheric spectra were computed by first using the code TLUSTY version 200(Hubeny 1988) to calculate the atmospheric structure and SYNSPEC version 48 (Hubeny and Lanz 1995) to construct synthetic spectra. We compiled a library of photospheric spectra covering the temperature range from 15,000K to 70,000K in increments of 1000 K, and a surface gravity range, log $g = 7.0 - 9.0$, in increments of 0.2 in log $g$.

We determined separately for each spectrum, the best-fitting single temperature white dwarf model and the best-fitting accretion disk-only model. Depending upon the success of these fits, we tested whether a combination of a disk plus a white dwarf significantly improved the fit.
Using two $\chi^{2}$ minimization routines, either IUEFIT for disks-alone and photospheres-alone or DISKFIT for combining disks and photospheres or two-temperature white dwarfs, $\chi^{2}$ values and a scale factor were computed for each model or combination of models. 
The scale factor, $S$, normalized to a kiloparsec and solar radius, can be related to the white dwarf radius R through: $F_{\lambda(obs)} = S H_{\lambda(model)}$, where $S=4\pi R^2 d^{-2}$, and $d$ is the  distance to the source. For the white dwarf radii, we use the mass-radius relation from the evolutionary model grid of Wood (1995) for C-O cores. The best-fitting model or combination of models was chosen based not only upon the minimum $\chi^{2}$ value achieved, but the goodness of fit of the continuum slope, the goodness of fit to the observed Lyman Alpha region and consistency of the scale factor-derived distance with other distance estimates in Table 1.

In Table 3, we list the best-fitting parameters of the three systems,
where the entries by row are (1) the system name, (2) white dwarf mass, (3) inclination angle, (4) best-fitting model distance in pc, (5) scale factor, (6) $\chi^{2}$ value and (7) $\dot{M}_{\sun}$ yr$^{-1}$).

\begin{deluxetable}{lrrc}
\tabletypesize{\scriptsize}

\tablecaption{Synthetic Spectral Fitting .}
\tablecolumns{4}
\tablewidth{0pt}
\tablehead{
\colhead{} & \colhead{V363 Aur} & \colhead{RZ Gruis} & \colhead{AC Cnc}
}
\startdata
$M_{wd}$    & $0.80M_{\sun}$      & $0.55M_{\sun}$      & $0.80M_{\sun}$ \\
Inclination & $75^{\circ}$        & $60^{\circ}$        & $75^{\circ}$  \\
Distance    & 134 pc              & 116 pc              & 105 pc \\ 
$\chi^{2}$  & 2.24                & 4.31                & 1.88    \\
$\dot{M}$   & 1$\times10^{-09}$ $M_{\odot}$/yr  & 1$\times10^{-08}$ $M_{\odot}$/yr  &  1$\times10^{-10}$ $M_{\odot}$/yr\\
\enddata
\end{deluxetable}

V363 Aur is normally observed at V=14.2 but has been observed to be as faint as V = 15. It has never been seen at a low brightness state thus suggesting its 
classification as a member of the UX UMa subclass. The FUV spectrum reveals a steeply rising continuum toward shorter wavelengths and a line spectrum dominated by strong emission features due to C IV and He II. The He II strength is strong for a non-magnetic nova-like, raising the possibility that the white dwarf could be magnetic. Absorption features due to Si III (1300), C II (1335) and O V (1371) are also seen along with the C IV and He II emission. On the basis of V363 Aur having an optically thick, steady state accretion disk, Hoare and Drew (1991) used the Zanstra method to determine a boundary layer temperature of 70000K to 100000K using both the He II 1640 flux and the He II 4686 flux.

For the model fitting to V363 Aur, we adopted a white dwarf mass M$\mathrm{wd} = 0.86$ $M_{\sun}$, inclination i = 73 degrees, and a reddening value of E(B-V)  = 0.3 (see Table 1). We carried out model accretion disk fitting using optically thick, steady state disk models. Our best-fitting accretion disk model to the de-reddened spectrum SWP25335 consisted of a white dwarf mass $M_{wd} = 0.8 M_{\odot}$ ($log(g)= 8.26$), an inclination i = 75 degrees, a $\chi^2 = 2.24$ and yielded an accretion rate $\dot{M} = 1 \times 10^{-9}M_{\odot}$/yr. This best-fit gave a distance of 134 pc for V363 Aur. This best-fitting accretion disk model is displayed in Fig.1.

%Given the possible evidence (strong emission at He II (1640) and He II (4686)) 
%that V363 Aur is magnetic, we explored the effect that a  magnetically truncated 
%accretion disk has on the FUV spectral modeling. For example, among a set of 
%truncated disks that we tried, the truncated accretion disk that gave the best 
%fit (the least $\chi^2$ value) to V363 Aur was truncated (an inner hole) at 
%two white dwarf radii and had 
%$log(\dot{M}) = -9.5 M_{\odot}$/yr.  This fit yielded a distance of 150 pc 
%and had a $\chi^2 = 2.37$. The truncated  disk model with a truncation radius 
%of two white dwarf radii is displayed in Fig.2. In short, the truncated disk 
%gives a better fit (lower $\chi^2$ value) than a full disk but the distance 
%is about half of what we found for the best fitting full disk with 
%$log(\dot{M}) = -9.0 M_{\odot}$/yr.

RZ Gru is typically observed at V=12.3 but has been observed as faint as 13.4. Thus, it is always in a high brightness state with excursions in brightness of a magnitude or less, also characteristic of the UX UMa subclass of nova-like variables. The IUE spectrum reveals a strong continuum steeply rising toward shorter wavelengths with deep absorption features due to C IV (blended 1548, 1550 components), Si IV (blended 1393, 1402 components), N V (blended 1238, 1242 components). The N V, Si IV and C IV resonance lines all have blue-shifted absorption and weak, sharp P Cygni absorption flanking the short wavelength side of the profiles. These features may manifest a hot, outflowing wind arising from the inner accretion disk and boundary layer between the star and disk.  The wind absorption at C IV and N V are very deep as expected for a UX UMa system viewed at low inclination where a larger disk surface area is emitting continuum photons which are absorbed by the outflowing wind.

Relatively little is known about RZ Gru. We adopted a reddening of 
E(B-V) = 0.03 (Bruch and Engel 1994) and adopted a nominal white 
dwarf mass of $0.6 M_{\odot}$ ($log(g)= 8$)  in the absence of any published may 
value. From Table 1, the published orbital inclination is 61 degrees. 
With these parameters, we carried out accretion disk model fits 
keeping the accretion rate as a free parameter. Our best fitting 
model accretion disk to SWP18138 is displayed in Fig.2.  The best-fit 
parameters are given in Table 3. The model accretion disk had 
i = 60 degrees,  $M_{wd} = 0.55 M_{\odot}$  ($log(g)= 7.71$),  
a $\chi^2 = 4.314$ and yielded a distance of 116 pc and 
accretion rate $\dot{M}   = 1 \times 10^{-9}M_{\odot}$/yr. 
This optically thick, steady state accretion disk accounts for essentially 100\% of the FUV flux. 

The observation of AC Cnc was obtained when its V-magnitude was 14.2, 
which is fainter by one magnitude than its normal brightness of 13.5. 
However, the system has been observed as faint as 15.4. Therefore 
the IUE spectrum was taken during a fainter high state and not during 
its deep low state. The IUE spectrum is dominated by a strong 
C IV (1548, 1550) emission feature and weak emission lines of 
N V (1238, 1242), Si IV (1393, 1402) and He II (1640). While the high inclination of AC Cnc reduces the brightness of the disk thus enhancing the possibility that the underlying white dwarf is exposed, the fact that the disk is viewed nearly edge-on raises the possibility that disk material and magnetic structures extending above and below the disk plane may complicate or obliterate the line spectrum and continuum exhibited by the exposed white dwarf's photosphere.

We have adopted a reddening value, E(B-V) = 0.055 for AC Cnc, that we have obtained from the GALEX archive. Using this value, we de-reddened the IUE spectrum and re-analyzed it with our grid of photosphere and accretion disk models. We carried out the model fitting in two ways: (1) taking the distance as a free parameter and (2) fixing the distance at values taken in the literature. 
Our model fitting of the de-reddened spectrum of AC Cnc, treating the distance 
as a free parameter, and with $i = 75$ degrees, and $M_{wd} = 0.8 M_{\odot}$, resulted in the best-fitting model to the de-reddened spectrum displayed in Fig.3. 
This model consists of an accretion disk with an accretion rate of 
$10^{-10} M_{\odot}$/yr, and a cool white dwarf contributing about 20\% of the FUV flux, and a distance of 105 pc.

With a fixed distance of 450 pc (Thoroughgood et al. 2004),  the best-fitting accretion disk model to the de-reddened spectrum with $i = 75$ degrees and 
$M_{wd} = 0.8 M_{\odot}$ yielded an accretion rate 
$log(\dot{M})=-9.0 M_{\odot}$/yr. However, the fit is unsatisfactory. 
When we fit the original spectrum without de-reddening but with the distance fixed at 450 pc, we obtained an accretion rate between 
$log(\dot{M})=-9.0$ and $-9.5 M_{\odot}$/yr, 
closer to the value derived by Puebla et al. (2007). 

\section{Discussion}

It is interesting to compare our results for V363 Aur, RZ Gru and AC Cnc
with  a statistical study  by Puebla, Diaz \& Hubeny et al. (2007), which utilized a multi-parametric optimization model fitting method, and explored how well current optically thick accretion disk models fit the FUV spectra of nova-likes and old novae in a sample of 33 nova-like and old novae. This study included an assessment of the contribution of an
assumed 40,000K white dwarf to the FUV spectra but did not include the white dwarf flux contribution in each system explicitly as a free parameter in the fitting. Rather, they  used a parameter defined as the ratio of the integrated spectral flux between 1500\AA\ and 3250\AA\ of the 40,000K white dwarf model and disk model to estimate the flux contribution
of the white dwarf. If this ratio was $<$ 0.1, then the system was defined as disk-dominated. Indeed, the white dwarf contribution is expected to be minimal in these systems because their disks are thick and luminous and because at high inclination the inner disk, boundary layer and white dwarf
should be significantly obscured by vertical structure in the disk.
These authors adopted distances in the literature while in our model fitting, we kept the distance, the white dwarf temperature and the orbital inclination as free parameters.

For V363 Aur, their range of parameters in their statistical fitting was an inclination range 60 to 80 degrees, white dwarf mass range of 0.7 to 
1.1 M$_{\sun}$ and an adopted distance of 550 pc for which their best fitting solution was an accretion disk with an accretion rate of 4.6$\times10^{-9}$ M$_{\sun}$ $yr^{-1}$. Their derived value of the accretion rate is a factor of ~5 larger than our derived accretion rate. We find that the major cause of the distance difference arises from the adoption of the reddening. Our adopted reddening, E(B-V) = 0.3 is taken from Szkody and Crosa (1981) and we consider it to be the most reliable value. Rutten et al. (1992) adopted E(B-V) = 0.1 following Schlegel et al. (1986). The difference in reddening itself generates a large difference in the distance. The galactic reddening is extremely large in the line of sight to V363 Aur, as large as 1.5, so if V363 Aur were very distant (say 500-1000 pc), then it would have an even larger reddening. 

For RZ Gru, Puebla et al.(2007) explored inclinations in the range of 10 
degrees to 35 degrees, white dwarf masses in the range 0.8 to 1.2 
$M_{\odot}$ and adopted a distance of 440 pc. 
Their best-fitting solution was an accretion disk with an accretion rate of 
$ 2 \times 10^{-9} M_{\sun}$yr$^{-1}$. Thus, their derived value of 
$\dot{M}$ is a factor of 5 smaller than the value that we derive for RZ Gru.  Stickland et al.(1984) derived a {\it i.e. upper limit} distance to RZ Gru of 440 pc. Moreover, their result used an exceedingly crude fit to the SED, 
ignoring the inclination angle of the disk. Their disk has 
$T_{max}=39,000$K, and an inner radius of 9,000 km. From the far more 
sophisticated published grid of Wade and Hubeny (1998) and from standard 
accretion disk theory, this corresponds to a WD mass of about 
$\sim 0.55 M_{\odot}$ (perhaps slightly smaller) and the corresponding 
disk (for that WD mass with a disk having $T_{max} = 39,000$K) has a mass 
accretion rate of $10^{-8.5} M_{\odot}$/yr. Such a disk with a low 
inclination gives a distance of 400 pc but gives a very bad fit 
(the slope does not fit properly). However, with an inclination of 
60 degrees, that same disk gives a slightly better fit and a distance 
of 230pc. We note that the best fit to the IUE SWP data is obtained 
assuming $i= 60$ deg with a mass accretion rate of $10^{-9.0} M_{\odot}$/yr for 
$M_{wd}=0.55 M_{\odot}$. This gives a distance of 117pc. 

We confirm the finding by Puebla et al. (2007) that the white dwarf component in AC Cnc is a non-negligible contributor to the FUV flux. However, the accretion rate that we obtained is nearly a factor of 10 lower than the rate determined by Puebla et al.(2007). We now have a reddening value, E(B-V) = 0.055 for AC Cnc, that we have obtained from the GALEX archive. Using this value, we de-reddened the IUE spectrum and re-analyzed it with our grid of photosphere and accretion disk models. The fit with the distance kept fixed at 450 pc is unsatisfactory. When we fit the original spectrum without de-reddening, we obtained an accretion rate between $log(\dot{M})  = -9.0$ and $-9.5 M_{\odot}$/yr,
closer to the value derived by Puebla et al. (2007). 

Our model-derived distances are considerably closer than the distances cited for AC Cnc and V363 Aur in the published literature. For example, for AC Cnc Shugarov (1981), Patterson (1984), Zhang (1987), Warner (1987) and Thoroughgood et al.(2004) obtained distances between 400 and 800 pc. For V363 Aur, Szkody \& Crosa (1981), Berriman (1987), Rutten et al. (1992) and Thoroughgood et al. (2004) obtained distances between
450 amd 1300 pc. Moreover, distances from an absolute magnitude calibration using infrared 2MASS magnitudes were derived by Ak et al. (2007). They found  704 +161/-208 pc for AC Cnc and 790 +187/-245 pc for V363 Aur. This IR calibration agrees with CV distances measured from trigonometric parallaxes (Gariety \& Ringwald 2011). Furthermore, the very similar nova-like BT Monocerotis (Porb=8.01 h) is measured to be at d=1700±300 pc (Smith, Dhillon \& Marsh 1998), although a closer distance of 972 +286/-406 pc is reported by Ak et al. (2007). However, since nearly all of the published distances were derived in some degree from assumed and measured properties of the Roche lobe-filling donor star, which is not only evolved but also irradiated (Thoroughgood et al. 2004), then it is possible that the uncertainties in the distances may be larger than what is estimated. Of the four distances in the literature, Patterson's (1984) distance is based upon a relationship between the $H_{alpha}$ equivalent width - $M_v$ relation plus "properties" of the  secondary (which is probably evolved) to obtain his distance. Given the difficulty in determining the correct spectral type and luminosity class (V or IV?) of the secondaries in AC Cnc and V363 Aur, we believe a shorter distance to the two objects cannot be definitively ruled out. 

It is obvious that the IUE spectra are not high quality data, especially at the shorter wavelengths of the IUE SWP spectra. The Lyman alpha region is strongly contaminated with airglow emission and the continuum level across the spectrum is difficult to identify
as there are absorption lines and emission lines seen across the spectrum. Some of these features in both eclipsing systems, may be due to magnetic structures or gas above the disk midplane. It is also possible that the continuum slope of the observed FUV spectra are affected by additional reddening (circumbinary?). Clearly, higher quality, phase-resolved FUV spectra are needed to confirm (or deny) our results. 

At their respective orbital periods, the secondaries of V363 Aur and AC Cnc should be G7 +/-2 and K2 +/- 1 respectively while the uncertainty in the orbital period of RZ Gru complicates a spectral type assignment. However, Stickland et al. (1984) estimated the secondary in RZ Gru to be an F5 dwarf. The secondaries in the three systems may be similar to each other in their degree of magnetic activity. Their convection zones and expected slower rotation may imply a lower level of magnetic activity. Since all three systems are nova-like variables of the UX UM type and have similar periods, their 
accretion rates should be similar. This suggests the possibility that the white dwarf temperature may be the key factor that differentiates the observed behavior and accretion rates of the three systems. Further investigations of the long period systems are clearly warranted.

It is a pleasure to acknowledge the support of this research by NSF grant AST08-07892 to Villanova University. Summer undergraduate research support was also provided by the NASA-Delaware Space Grant Consortium. PG is thankful to William P. Blair for his kind hospitality at the Henry Augustus Rowland Department of Physics \& Astronomy at the
Johns Hopkins University, Baltimore, MD.

\begin{figure} 
\epsscale{.75}
\plotone{f1.eps}
\caption{Plot of flux versus wavelength for the spectrum SWP35335 
of the UX UMa system V363 Aur. The solid curve is the best-fitting 
accretion disk model (see text for details).
}
\end{figure}

\begin{figure}
\epsscale{.75} 
\plotone{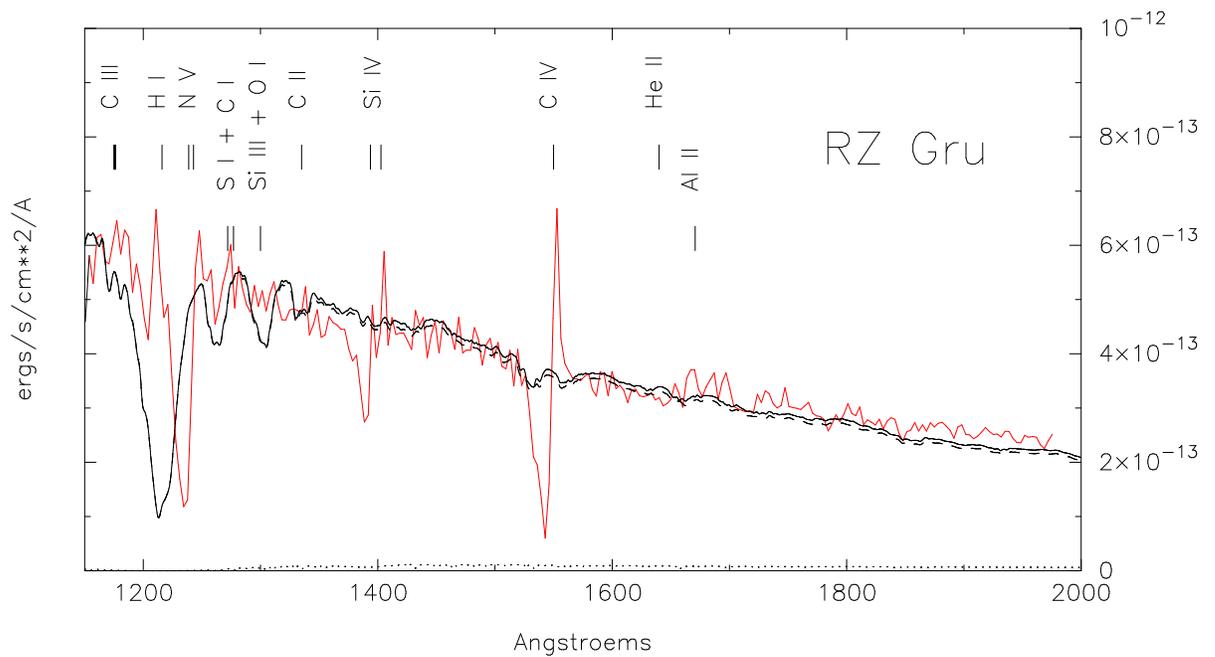}
\caption{The flux versus wavelength plot for the spectrum SWP18138 of RZ Gru. 
The solid curve is the best-fitting
accretion disk model (see text for details).
}
\end{figure}

\begin{figure}
\epsscale{.90} 
\plotone{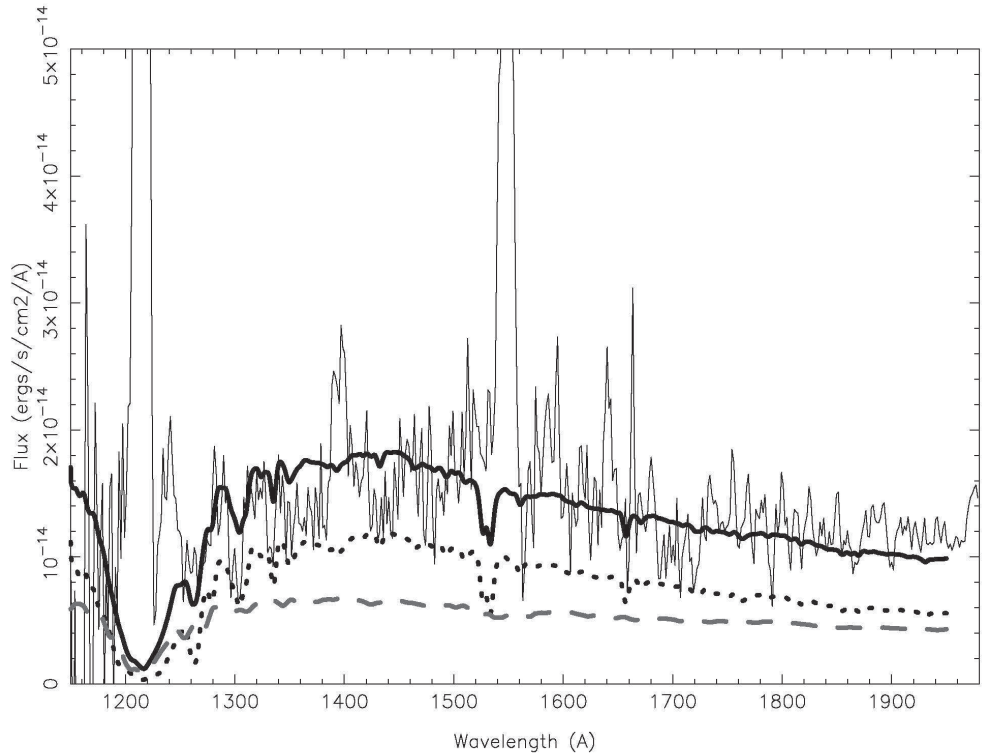}
\caption{The best-fitting combination accretion disk + white dwarf photosphere
synthetic fluxes to the spectrum SWP18734 of AC Cnc in its constant 
high brightness state (for E(B-V) = 0.055).  The top solid curve is the
best-fitting combination, the dotted curve is the contribution of the white dwarf 
alone and the dashed curve is the accretion disk synthetic spectrum 
alone (see text for details).
}
\end{figure}

\end{document}